\documentclass[journal=jpcafh,layout=twocolumn,manuscript=article]{achemso}

\usepackage{graphicx,color}
\usepackage{amsmath,amsfonts,amssymb}
\usepackage{dcolumn}

\newcommand{\fig}{Fig.}
\newcommand{\Fig}{Fig.}
\newcommand{\figref}[1]{\fig~\ref{#1}}
\newcommand{\Figref}[1]{\Fig~\ref{#1}}

\newcommand{\tabref}[1]{table~\ref{#1}}

\renewcommand{\eqref}[1]{equation~(\ref{#1})}

\usepackage{xcolor}
\usepackage{tikz}
\usetikzlibrary{shapes.geometric}
\newcommand\mrahmen[3][]{%
  \tikz[anchor=base,baseline]\node[inner sep=3pt,draw=#2,#1]{$\displaystyle#3\mathstrut$};}
\colorlet{mfarbe}{blue}

\renewcommand{\vec}[1]{\boldsymbol{#1}}
\newcommand{\vecnp}[1]{\mathbf{#1}}

\newcommand{\new}[1]{#1} 

\makeatletter
\def\@dotsep{4.5}
\makeatother
\title{A Dual-Level Approach to Instanton Theory}
\author{Jan Meisner}
\author{Johannes K\"{a}stner}
\email{kaestner@theochem.uni-stuttgart.de}
\affiliation{Institute for Theoretical Chemistry, University of Stuttgart, Pfaffenwaldring 55, 70569 Stuttgart,Germany}

\date{}

\begin{document}

\begin{abstract}
Instanton theory is an established method to calculate rate constants of chemical reactions including atom tunneling.
Technical and methodological improvements increased its applicability.
Still, a large number of energy and gradient calculations is necessary to optimize the instanton tunneling path and
2$^{\text{nd}}$ derivatives of the potential energy along the tunneling path have to be evaluated,
restricting the  range of suitable electronic structure methods.
To enhance the applicability of instanton theory, we present a dual-level approach 
in which instanton optimizations and Hessian calculations are performed using an
efficient but approximate electronic structure method and the 
potential energy along the tunneling path is recalculated using a more
accurate method.
This procedure extends the applicability of instanton theory to
high-level electronic structure methods for which analytic gradients
may not be available, like local linear-scaling approaches.
%
We demonstrate for the analytical Eckart barrier and three molecular systems how the dual-level instanton approach corrects for  the largest part of the error caused by the inaccuracy of the efficient electronic structure method.
This reduces the error of the calculated rate constants significantly.

\end{abstract}

\maketitle


\section{Introduction}

The accurate description of atom tunneling is crucial for the calculation of
precise reaction rate constants.  An efficient yet accurate computational
method to incorporate atom tunneling is semiclassical instanton
theory.\cite{langer1967,miller1975,coleman1977,callan1977,gildener1977}
Instanton theory can be understood as a quantum mechanical analog to
transition state theory (TST), thus, referred to as harmonic quantum
transition state theory.\cite{mills1998,andersson2009} The idea
of instanton theory is to include nuclear quantum effects by statistical 
Feynman path integrals. Partition functions are approximated by the steepest
descent approach, i.e.  optimizing the most likely tunneling path, the
so-called instanton, which is a 1$^{\text{st}}$~order saddle-point of the
Euclidean action $S_{\text{E}}$, and taking fluctuations into account within
the harmonic 
approximation.\cite{ric09,rommel2011,rommel2011b,
  richardson2016a,richardson2016b,mcconnell2017,mcc17a} It is a successful
compromise between accuracy and computational efficiency as it does not
require a global potential energy surface, but can be used in combination with
on-the-fly electronic structure
calculations.\cite{kaestner2014,meisnerreview2016}

The applicability of instanton theory to the quantification of atom
tunneling in molecular systems has been frequently demonstrated in the
past
decade.\cite{mills1994,mills1997,goumans2010,jonsson2011,goumans2011,goumans2011a,meisner2011,alvarez-barcia2014,kryvohuz2014,richardson2016,beyer2016,song2016,meisnerreview2016,lamberts2017,lamberts2017b,son17}
While traditionally instanton paths were often guessed or approximated
using further
assumptions,\cite{siebrand1999,smedarchina1997,tautermann2002,tautermann2002b}
a modified Newton--Raphson (NR) instanton optimizer\cite{rommel2011}
allows efficient instanton search for molecular systems with many
degrees of
freedom.\cite{rommel2011,goumans2011a,meisner2011,rommel2012,alvarez-barcia2014,
  kryvohuz2014,richardson2016,beyer2016,song2016,lamberts2017,lamberts2017b,son17}
%

In practical applications, the instanton path $\vecnp{y}_{\text{inst}}$ is
discretized into $P$ replicas or images. From these, the Euclidean action
$S_\text{E}=S_0/2+S_\text{pot}$ can
be calculated with
\begin{equation}
S_0 = 
\frac{P}{\beta\hbar}
 \sum_{k=1}^{P}
(\vec{y}_{k}
-\vec{y}_{k-1}
)^2,
\label{eq:Snull}
\end{equation}
and
\begin{equation}
S_{\text{pot}}=
\frac{\beta\hbar}{P}
\sum_{k=1}^{P}
 V(\vec{y}_k).
\label{eq:Spot}
\end{equation}
Here, $\vec{y}_k$ are the $N$ mass-weighted coordinates of the $k^{\text{th}}$ image, $\hbar$ is
the reduced Planck's constant, $\beta = (k_{\text{B}} T)^{-1}$, $T$ is the
temperature, and $k_{\text{B}}$ is Boltzmann's constant. The reduced action
$S_0$ is a measure of the path length and the distribution of the images along the path, while $S_{\text{pot}}$ basically
averages the potential along the instanton path. The potential $V(\vec{y}_k)$
at each image $k$ is obtained from electronic structure calculations. 

Instanton theory is, in its standard form, applicable only below a
system-specific crossover temperature
\begin{equation}
T_\text{c}=\frac{\hbar\omega_\text{b}}{2\pi k_\text{B}}
\end{equation}
which depends on the absolute value of the imaginary frequency of the
transition state $\omega_\text{b}$.  Methods to extend it to above
$T_\text{c}$ are available,\cite{zha14,richardson2016b,mcconnell2017,mcc17a} but
will not be used here.

The instanton rate constant is obtained by
\begin{multline}
k_{\text{inst}}= 
\sqrt{\frac{S_0 P}{2 \pi \beta \hbar ^2} } \;
\sqrt{
\frac{\prod_{l=1}^{NP} \lambda_l^{\text{RS}}}
     {\prod_{l=1}^{'NP} | \lambda_l^{\text{inst}}| }
}\times
\\
\exp\left(-\frac{S_\text{E}(\text{inst})-S_\text{E}(\text{RS})}{\hbar}\right).
\label{eq:instantonrate}
\end{multline}
Here, $S_\text{E}(\text{inst})$ refers to the Euclidean action of the
instanton path, while $S_\text{E}(\text{RS})$ is the Euclidean action of the
reactant state. In the latter, the optimal Feynman path is collapsed to the
minimum on the potential energy surface, i.e. $S_0(\text{RS})=0$ and
$S_\text{E}(\text{RS})=\beta\hbar V(\vec{y}_\text{RS})$. The eigenvalues
$\lambda_l^{\text{RS}}$ and $\lambda_l^{\text{inst}}$ refer to the second
derivatives of $S_\text{E}$ with respect to all coordinates of all images.  To
evaluate the latter, Hessians of $V(\vec{y}_k)$ have to be calculated for all
images along the instanton path. One eigenvalue $\lambda_l^{\text{inst}}$ is
negative, seven are zero (six for linear molecules), all others are
positive.\cite{kaestner2014} The zero-eigenvalues are left out of the product,
thus the prime in the product sign in \eqref{eq:instantonrate}. Together with
the instanton optimization, the calculation of the Hessians is the most
time-consuming step in computational applications of instanton theory.
In one example described below, a sixteen-atomic molecular system
(i.e. $N=48$), the instanton is discretized into $P=100$ images.  This leads
to a 2400-dimensional optimization problem (because the instanton, a closed
Feynman path, covers the same line in configuration space twice, forward and
backward).  Even though the instanton optimization often requires just a few
iterations,\cite{rommel2011} for every iteration $P/2=50$ energy and gradient
evaluations are required.  For the rate constant calculation, 50 Hessian
calculations have to be carried out. These computations provide the rate
constant for one value of the temperature. It is obvious that instanton
calculations involve a significant computational effort.

In this paper, a dual-level approach to instanton theory is presented where
for the time-consuming calculations a fast approximative method is applied to
calculate $V(\vec{y})$.  Subsequent energy calculations with a more accurate
electronic structure method improve the quality of the calculated rate
constant significantly.

Three different molecular systems were used to 
illustrate the performance of the dual-level instanton approach:
the isomerization of HNC to HCN, 
an intramolecular [1,5] hydride shift,
and 
the bimolecular hydrogen atom transfer reaction of
NH$_2$ + H$_2 \rightarrow$ NH$_3$ + H.
For each system,
reaction rate constants were calculated using the dual-level
approach and are compared to the results of the conventional instanton theory
at both the basic and accurate electronic potential.
%

\section{Background and Computational Details}
In computational chemistry, the 
evaluation of the electronic potential energy
is often the most time-consuming step.
One possible compromise to obtain reliable results within a reasonable time
is a dual-level approach. An approximate, but less demanding
computational method is used 
for geometry optimizations and Hessian calculations in what will be
referred to as the basic potential in the following.
Subsequently, 
the potential energy is calculated using a more accurate, but more time
consuming electronic
structure method, which provides the refined potential.
The fundamental assumption is that the basic potential is still able to
reproduce the molecular geometries of the stationary points sufficiently accurate. 
Thermochemical results obtained in this way generally agree very well
with results obtained with the refined potential. Also some calculations of rate constants using different tunneling corrections such as SCT and $\mu$OMT utilized a dual-level approach successfully.\cite{corchado1998,sekusak1999,huang2001}
Dual-level or even multi-level approaches have been used for the calculation of potential energy surfaces necessary for the computation of vibrational spectroscopy.\cite{pflueger2005}

Here, we show that the dual-level instanton approach 
leads to excellent results in terms of accuracy and computational effort.
First the discretized instanton is optimized on the  basic
potential. Hessians of the potential energy are also calculated using
the basic potential, providing $\lambda_l^{\text{RS}}$ and $\lambda_l^{\text{inst}}$.
Subsequently the energies at all images, $V(\vec{y}_k)$, as well as $V(\vec{y}_\text{RS})$ are
calculated on the refined potential. Note, that no geometry
optimizations at the refined potential are necessary, just energy
calculations. From these, $S_{\text{pot}}$ is obtained using the
refined potential. 
This is assumed to correct the error caused by the inaccuracy of the
basic potential.
The working hypothesis is that the basic potential is able to reproduce the shape and image distribution of the instantons qualitatively well.
The dual-level instanton rate constant reads 
\begin{multline}
k_{\text{inst}}^{\text{\textbf{\textcolor{blue}{Dual}}}}= 
\sqrt{\frac{S_0 P}{2 \pi \beta \hbar ^2} } \;
\sqrt{
\frac{\prod_{l=1}^{NP} \lambda_l^{\text{RS}}}
     {\prod_{l=1}^{'NP} | \lambda_l^{\text{inst}}| }
}\times
\\
\exp\left(
\frac{
- \frac{ S_0(\text{inst})}{2}
- \mrahmen[rounded corners]{mfarbe}{\textcolor{blue}{\boldsymbol{S}}_{\text{\textcolor{blue}{\textbf{pot}}}}}({\text{inst}})
+ \mrahmen[rounded corners]{mfarbe}{\textcolor{blue}{\boldsymbol{S}}_{\text{\textcolor{blue}{\textbf{pot}}}}}({\text{RS}})
}{\hbar}
\right)
\label{eq:instantondual-level}
\end{multline}
where highlighting denotes the refined potential.

Hitherto, different studies combined approximate instanton methods with
dual-level approaches\cite{smedarchina1997,tautermann2002,smedarchina2012} \new{or
used basic-potential instantons as initial guess for the optimization of
high-potential ones to obtain tunneling splittings.\cite{mil03a,mil04}}
Only an instanton path fully optimized on the same potential energy surface as
the one used for the Hessian calculations ensures the proper eigenvalue
structure of $\lambda_l^{\text{inst}}$.
%
Thus, in this work, instantons are optimized in all dimensions using the basic
potential, \emph{i.e.,} they are true 1$^{\text{st}}$-order saddle points of
the Euclidean action $S_{\text{E}}$. The refined potential is then used to
correct $S_{\text{pot}}$.
%
%

For dual-level approaches in quantum chemistry, the nomenclature
\begin{center}
Method$_A$/basis set$_A$ /\!/ Method$_B$/basis set$_B$
\end{center}
is used. Here, $A$ labels the level of theory and basis set of the refined
potential and $B$ refers to the basic potential.
For simplicity, the basis set declaration is omitted in the shorthand notation
\begin{center}
Method$_A$ /\!/ Method$_B$
\end{center}
when possible.

In this work all geometry optimizations, including intrinsic reaction
coordinates (IRCs) and instantons, as
well as rate calculations were performed using the DL-FIND optimization
library\cite{ kaestner2009} interfaced to
Chemshell.\cite{sherwood2003,metz2014} 
IRCs were optimized using a Hessian-predictor-corrector algorithm.\cite{meisner2017b}
Instantons were optimized until the
maximum component of the gradient is smaller than $1\cdot 10^{-8}$ atomic
units (scaled relative to the electron's mass) using the adapted,
quadratically convergent Newton--Raphson
algorithm.\cite{rommel2011,rommel2011b}
CCSD(T)-F12\new{a}\cite{adler2007,adler2009} calculations were performed using
molpro\cite{werner2012} with default settings and the cc-pVDZ-F12 basis set
throughout.\cite{peterson2008} All DFT calculations were performed with
Turbomole version 7.0.1\cite{turbomole}.  SCF energies were converged to an
accuracy of $10^{-9}$~Hartree on an \emph{m5} multi-grid.\cite{eichkorn1997}

For the Eckart barrier and reaction 1, $P=200$ images were used, for reaction
2, $P=100$ was used for all temperatures.  For reaction 3, $P=154$ images were
used for the DFT instanton calculations.  For the instantons calculated on
CCSD(T)-F12 level, 40 images were used down to 219~K, 78 images down to 131~K,
and 154 and 306 images for 119~K and 109~K, respectively. Tests showed, that
the rate constants were converged with respect to the number of images $P$
within the comparisons done here.

This paper demonstrates the applicability of the dual-level instanton method.
Here, the focus lies on the comparison of the dual-level instanton method with 
conventional instanton theory. As absolute values of the reaction rate
constants are of less importance for this study, no comparison with
literature values is done.

\section{Results}

\subsection{The Eckart barrier}

As a first example, we apply the dual-level scheme to the analytic
one-dimensional Eckart\cite{eck30} barrier in order to distinguish specific
inaccuracies of the basic potential and their effects on the dual-level
approach. We investigated several different scenarios of which we show one
typical case and the one where we found the weakest performance of the
dual-level approach in order to learn about the limits of the method.

The potential of the Eckart barrier is given by
\begin{equation}
V(x)=\frac{V_\text{A}y}{1+y} + \frac{V_\text{B}y}{(1+y)^2},  
\end{equation}
\begin{equation}
y= \frac{V_\text{A}+V_\text{B}}{V_\text{B}-V_\text{A}}
   \exp(\alpha x).
\end{equation}
Here $V_\text{A}$ determines the exothermicity: while for large negative
reaction coordinates $x$ the potential becomes zero, it approaches
$V_\text{A}$ for large positive $x$. The maximum of $V(x)$ is at $x=0$ and has
the value
\begin{equation}
V_\text{max}=\frac{(V_\text{A}+V_\text{B})^2}{4V_\text{B}}.
\end{equation}
For the refined potential we chose $V_\text{A}=-0.01$ Hartree
($-$26.3~kJ~mol$^{-1}$) and $V_\text{max}=0.01$ Hartree. These two parameters are
varied in the following to describe different basic potentials. The width
$\alpha$ is adjusted in all cases such that $T_\text{c}=300$~K.

\begin{figure}[htbp!]
    \includegraphics[width=8cm]{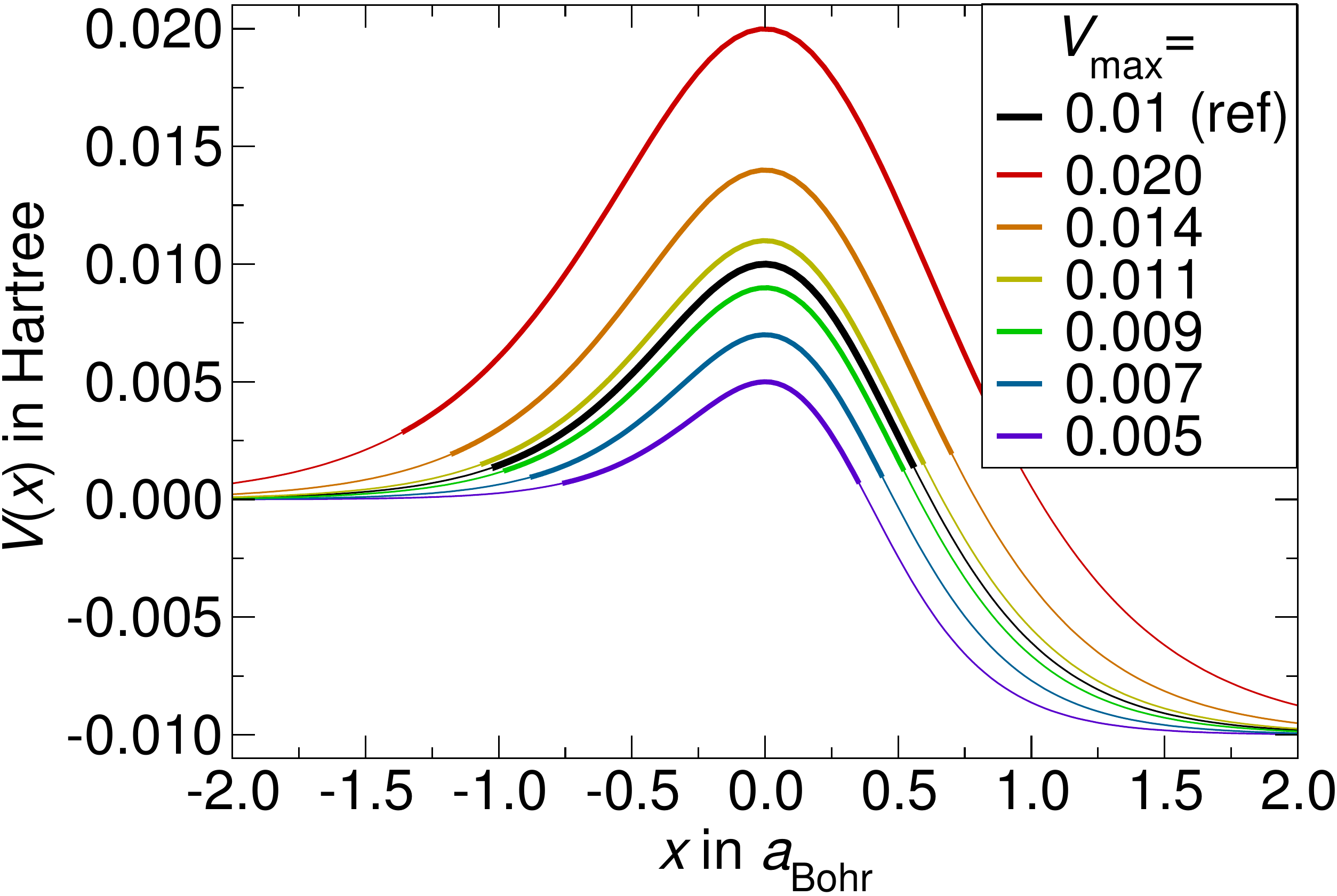}\\
    \includegraphics[width=7.5cm]{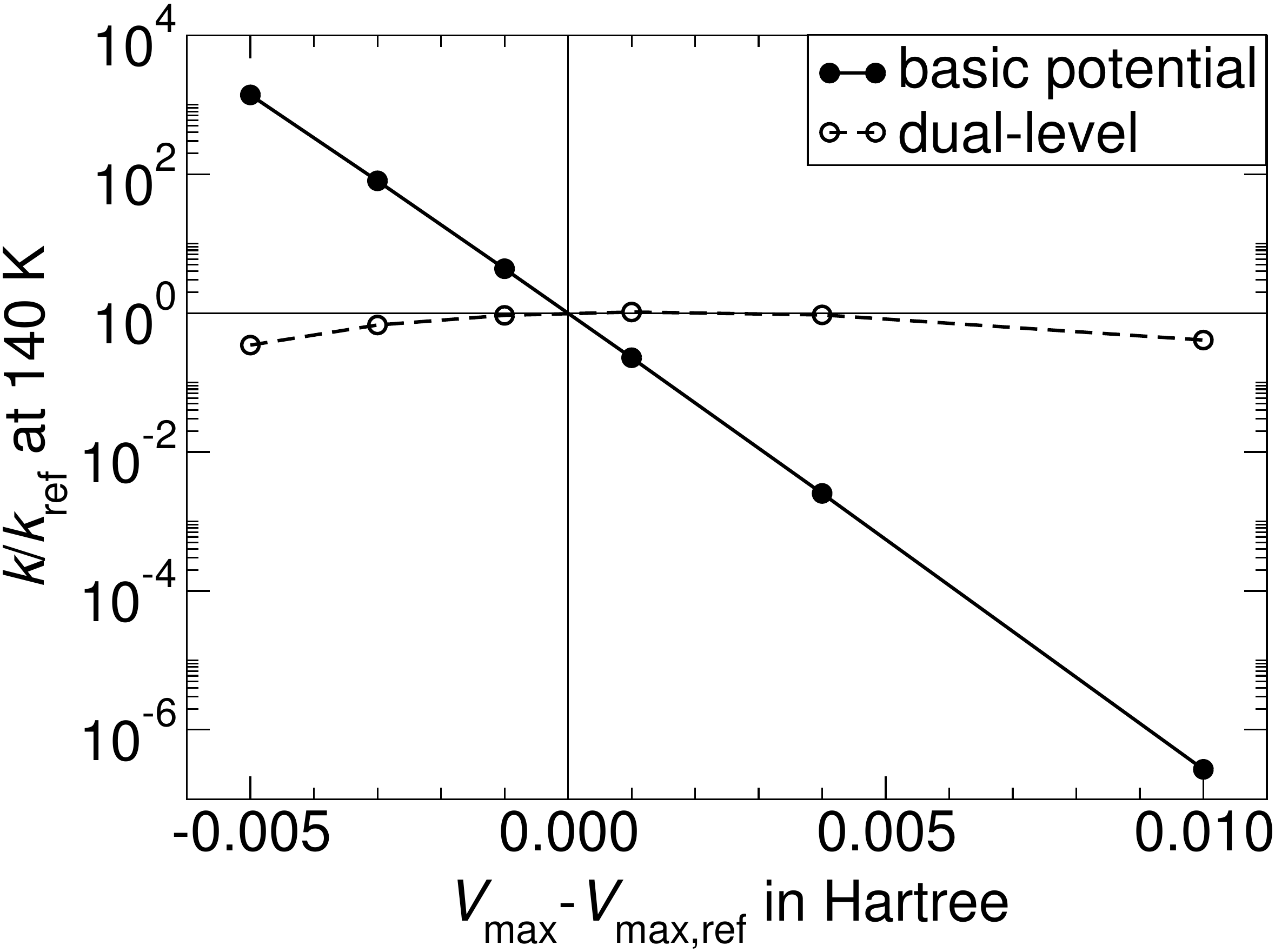}
  \caption{Effects of errors in the barrier height on the rate constants. Top:
    Barrier shapes for different values of $V_\text{max}$, thicker
    lines indicate instantons at 140~K. Bottom: Deviation of the rate constant as function of the
    deviation of the barrier height.
    \label{fig:dual:vmax}
  }
\end{figure}

The effect of an inaccurate description of the barrier height by the basic
potential is illustrated in \figref{fig:dual:vmax}. If $V_\text{max}$ is
varied from twice to half the reference value while keeping the reaction
energy and $T_\text{c}$ (and, thus, $\omega_\text{b}$) constant, the resulting
rate constant at 140~K increases by 3 or decreases by almost 7 orders of
magnitude. The dual-level scheme (hollow symbols in the lower graph of
\figref{fig:dual:vmax}) corrects for that very accurately, with remaining
errors of less than a factor of 3. A wrong barrier height is the typical error
of an approximate electronic structure method.

\begin{figure}[htbp!]
    \includegraphics[width=8cm]{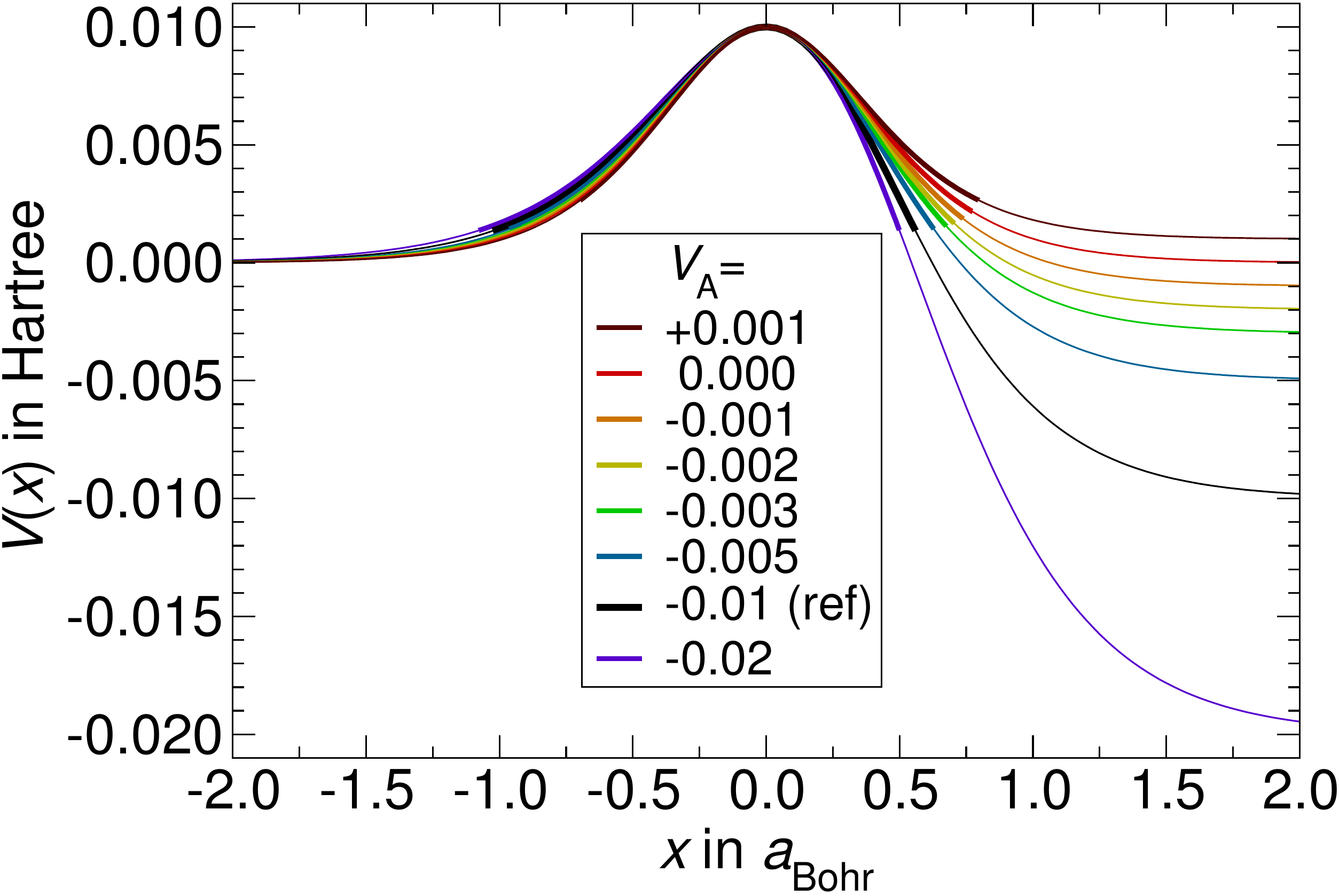}\\
    \includegraphics[width=7.3cm]{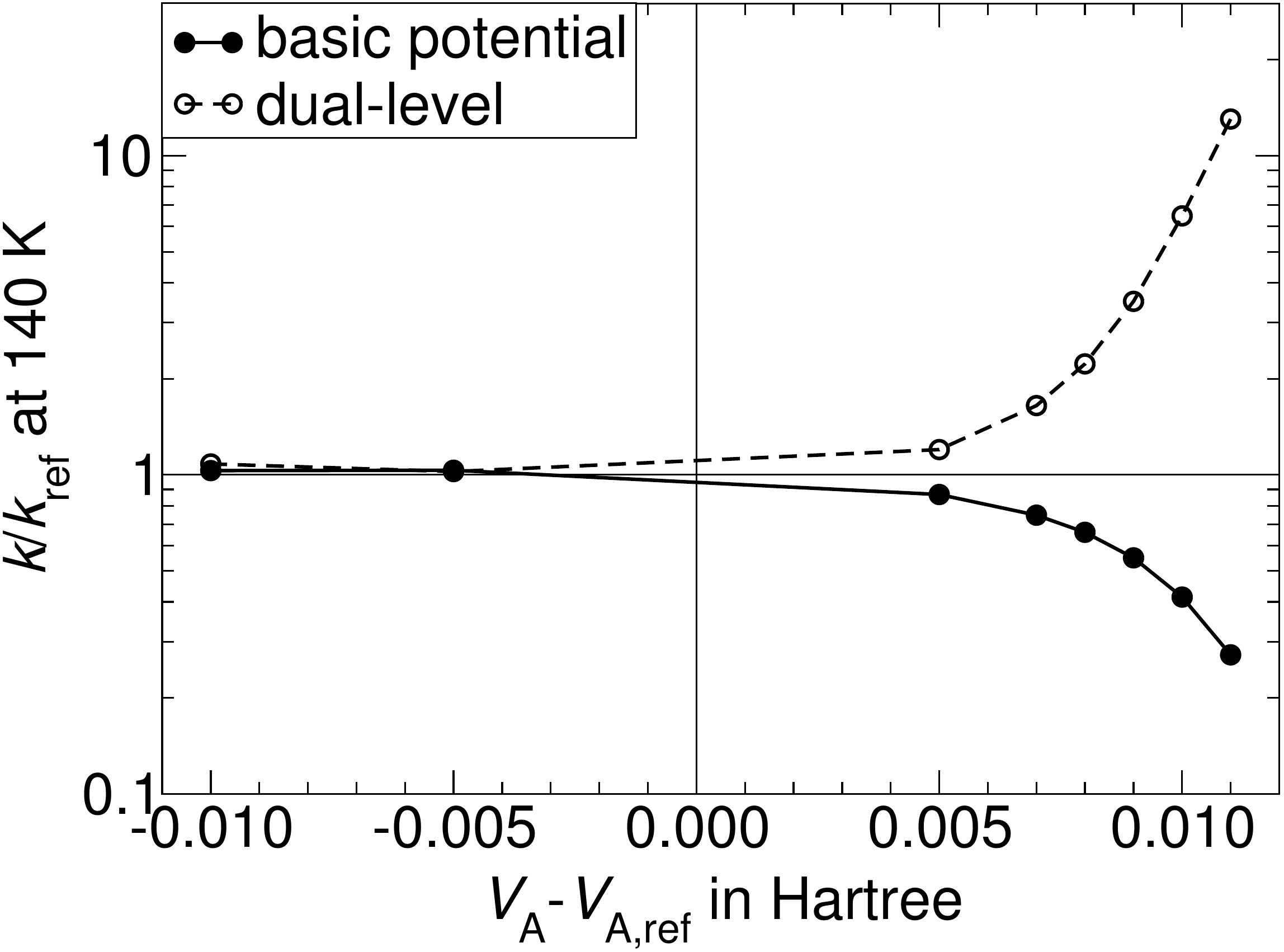}
  \caption{Effects of errors in the exothermicity ($-V_\text{A}$) on the rate constants. Top:
   Barrier shapes for different values of $V_\text{A}$, thicker
    lines indicate instantons at 140~K.
Bottom: Deviation of the rate constant as function of the
    deviation of the exothermicity.
    \label{fig:dual:va}
  }
\end{figure}

By contrast, the case where the barrier height is preserved, but the
exothermicity is described wrongly is shown in \figref{fig:dual:va}. Since the
rate constant primarily depends on the barrier height rather than the shape of
the barrier at the product side, the changes caused in the rate constant are
smaller. Especially if the basic potential predicts a too exothermic reaction,
the influence on the rate constant is negligible.  However, when the basic
potential describes the reaction energy to be too positive, we find the
extreme case where the dual-level approach fails, which can be seen from the
lower graph of \figref{fig:dual:va}. Whenever the exothermicity is too low
($V_\text{A}-V_\text{A,ref}>0$) the rate constant obtained by the dual-level
scheme is less accurate than the non-corrected basic rate constants. The
reason is that with a potential describing the reaction to be too  endothermic, the distribution of the
images is shifted more to the product side, which can not be corrected by the
dual-level approach. The refined potential energy of these images on the
product side is then erroneously low, which leads to an underestimation of
$S_\text{E}$ and an overestimation of the rate constant. \new{The
  cause of this difference is the change in the potential energy in
  the region of the instanton, rather than in the product well.} It should be noted
that this is the most extreme case we found, and still the overall error is
rather small.

\begin{table*}[htbp!]
  \caption{
    Potential energy barriers $V_A$, potential reaction energies $\Delta V$ and the
    respective values corrected by zero-point energy, $E_A$, and $\Delta E$
    for reaction 1: HNC~$\rightarrow$~HCN.
	Energies in kJ~mol$^{-1}$, 
        $T_{\text{c}}$ 
	in K. 
	\label{tab:dual-level:hnc}
      }
        \begin{tabular}{lrrr}
\hline
			&Basic Potential&	Refined potential&       Dual-level	\\ \hline
	Method		&B3LYP		&	CCSD(T)-F12     &         		\\
	Basis set	&def2-SVP	&	cc-pVDZ-F12	&              	\\
        $V_A$ 		&142.8		&	136.2		&	136.2		\\
        $\Delta V$	&$-57.2$	&	$-62.6$		&	$-62.4$		\\
        $E_A	$     	&128.9		&	123.5		&	122.4		\\
        $\Delta E$	&$-60.8$	&	$-65.5$		&	$-66.0$		\\
        $T_\text{c}$	&257.0		&	272.9		&			\\
\hline
\end{tabular}
\end{table*}

\subsection{Reaction 1: The Isomerization HNC $\rightarrow$ HCN}

The isomerization of HNC to HCN is a unimolecular prototype reaction with
well-defined reactant state structure (HNC) and product state structure (HCN).
Furthermore, the three-atomic system is small enough to carry out 
full conventional instanton calculations on the 
CCSD(T)-F12/cc-pVDZ-F12\cite{adler2007,peterson2008,knizia2009} level,
which was used as the refined potential.
For the basic potential,
B3LYP\cite{dirac1929,slater1951,vosko1980,bec88,lee88,bec93} was used in
combination with the def2-SVP basis set.\cite{weigend2005}

The reaction energies $\Delta V$ of basic and refined potential deviate by
5.4~kJ~mol$^{-1}$, see \tabref{tab:dual-level:hnc}.  The potential activation
barrier $V_A$ of B3LYP is higher than the CCSD(T)-F12 value by only
6.6~kJ~mol$^{-1}$.  The crossover temperatures are also quite similar
and deviate by just 15.9~K, that is $\approx$~6~\%.

\begin{figure}[htbp!]
    \includegraphics[width=8cm]{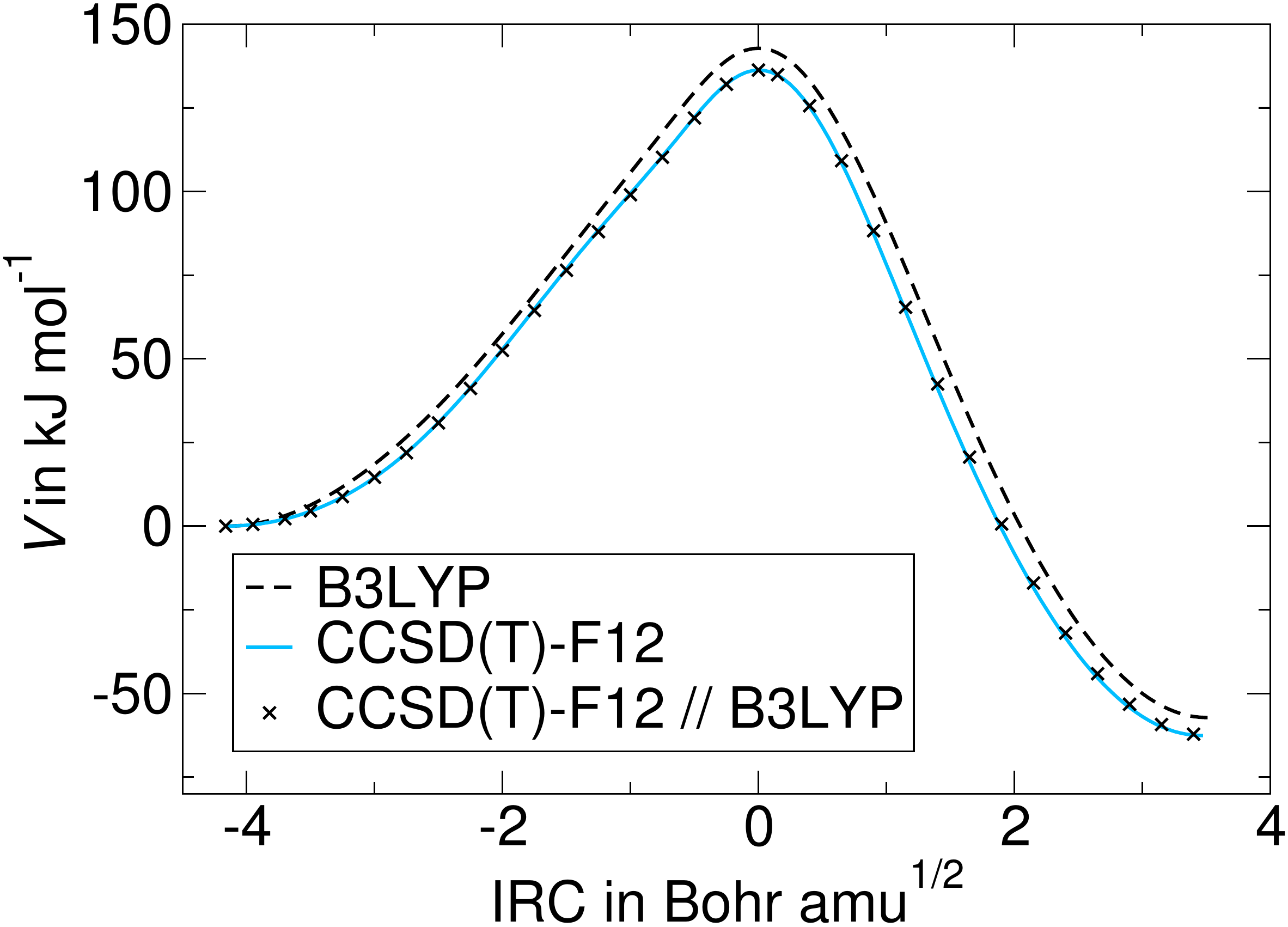}
  \caption{
    Potential energy along the IRCs of reaction 1. 
    \label{fig:dual-level:hncIRC}
  }
\end{figure}

The IRCs were optimized for both potentials, see
\figref{fig:dual-level:hncIRC}.  Additionally, CCSD(T)-F12 energy calculations
were performed on the IRC optimized with B3LYP, shown as blue crosses in
\figref{fig:dual-level:hncIRC}.  That potential energy curve coincides with
the potential energy along the IRC of CCSD(T)-F12, indicating that the IRC
geometries are similar.

\begin{figure}
    \includegraphics[width=8cm]{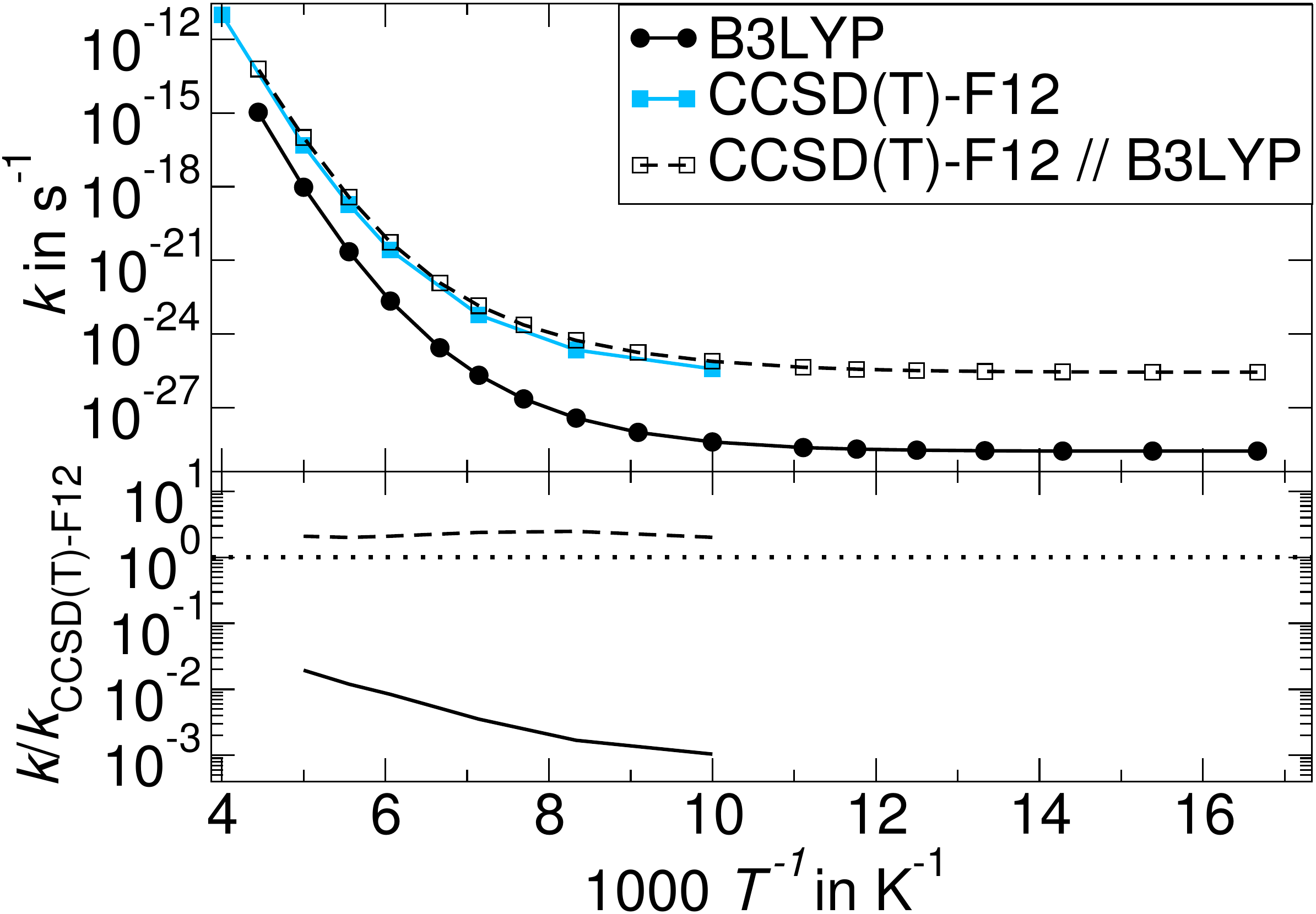}
  \caption{Top: Rate constants for reaction 1 with basic and refined
    potential as well as the dual-level method. Bottom: relative error of the
    rate constants with the basic potential and the dual-level method.
The dotted line denotes $k = k_{\text{CCSD(T)-F12}}$, i.e. no
error. 
    \label{fig:dual-level:hnc}
  }
\end{figure}

Rate constants for reaction 1 are shown in \Figref{fig:dual-level:hnc} as
Arrhenius plots, i.e. the logarithm of the rate constant is plotted against
$1/T$. Rate constants were obtained by conventional instanton theory using
both electronic potentials as well as the CCSD(T)-F12 /\!/ B3LYP dual-level
method.  For B3LYP and the dual-level method, reaction rate constants down to
60~K were calculated, for CCSD(T)-F12 down to only 100~K.  Due to the higher
activation barrier, the B3LYP reaction rate constants are lower than the
CCSD(T)-F12 reaction rate constants by a factor of 51.4 at 200~K and by a
factor of 961.1 at 100~K. The dual-level corrects that error to factors between
2.0 and 2.5, see \figref{fig:dual-level:hnc} and, thus, leads to excellent
agreement with the results obtained with full calculations on the refined
potential.

\new{Using reaction 1 as an example we want to point out the savings in computational
time: The optimization of one instanton and the subsequent rate
calculation at CCSD(T)-F12a level require 19,744 and 30,449 seconds,
respectively. On B3LYP level 626 and 1,305 seconds are required
for the instanton optimization and rate calculation on the same
computer infrastructure. The additional CCSD(T)-F12a energy
calculations for the dual-level approach require 323
seconds. Derivatives on the CCSD(T)-F12a level were obtained from
finite differences of energies, while DFT gradients were calculated
analytically. Thus, in this comparison, the dual-level approach saves
a factor of $>20$ in computational time.}


\subsection{Reaction 2: Intramolecular [1,5]-H-Shift}

\begin{figure}[b]
    \includegraphics[width=8cm]{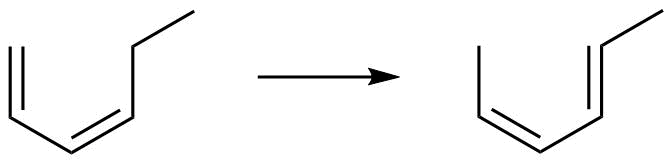}
  \caption{
        The [1,5] H shift in reaction 2.
	\label{fig:dual-level:sigmatropic}
  }
\end{figure}

In sigmatropic rearrangements, tunneling was observed in many cases.
Suprafacial [1,5] sigmatropic rearrangements were studied exhaustively, using
derivatives of 1,3(Z)-pentadiene.\cite{dormans1986,liu1993a} Although it was
initially unclear whether or not atom tunneling plays a crucial role in these
reactions,\cite{doering2006,doering2007} various studies have confirmed its
involvement.\cite{liu1993a,vanivcek2007,shelton2007,peles2008,zimmermann2010,kryvohuz2012,kryvohuz2014,meisnerreview2016}
The [1,5] sigmatropic rearrangement of 1,3(Z)-hexadiene to
2(E),4(Z)-hexadiene, \figref{fig:dual-level:sigmatropic}, is therefore an
appropriate test system for the dual-level instanton method.

Two density functionals, the BP86 GGA
functional\cite{dirac1929,slater1951,vosko1980,bec88,perdew1986} and
the BHLYP hybrid
functional,\cite{dirac1929,slater1951,vosko1980,bec88,lee88,becke1993}
both with the 6-31G* basis set,\cite{hariharan1973} were applied to obtain
different electronic potentials.

\begin{table*}[htbp!]
 \caption{
        Potential energy barriers $V_A$, potential reaction energies $\Delta V$ and the
        respective values corrected by zero-point energy, $E_A$ and $\Delta E$
        for reaction 2.
        Energies are in kJ~mol$^{-1}$, crossover temperatures $T_{\text{c}}$ in K.
        \label{tab:dual-level:sigmatropic}
	}
        \begin{tabular}{lrrrr}
\hline
        Method          &BP86		& BHLYP		& BP86 /\!/ BHLYP &  BHLYP /\!/ BP86			\\\hline
        Basis set       &6-31G*         &       6-31G*  &	      &			  \\
        $V_A$           &110.6		&	167.4	&109.8        &	166.3		  \\
        $\Delta V$      &$-18.0$	&	$-15.2$	&$-18.1$      &	$-15.3$		  \\
        $E_A    $       &100.3		&	156.4	&98.9         &	155.9		  \\
        $\Delta E$      &$-18.9$	&	$-16.4$	&$-19.3$      &	$-16.3$		  \\
        $T_\text{c}$    &308.8		&	394.7	&             &			  \\
\hline
\end{tabular}
\end{table*}

\begin{figure}[h!]
    \includegraphics[width=8cm]{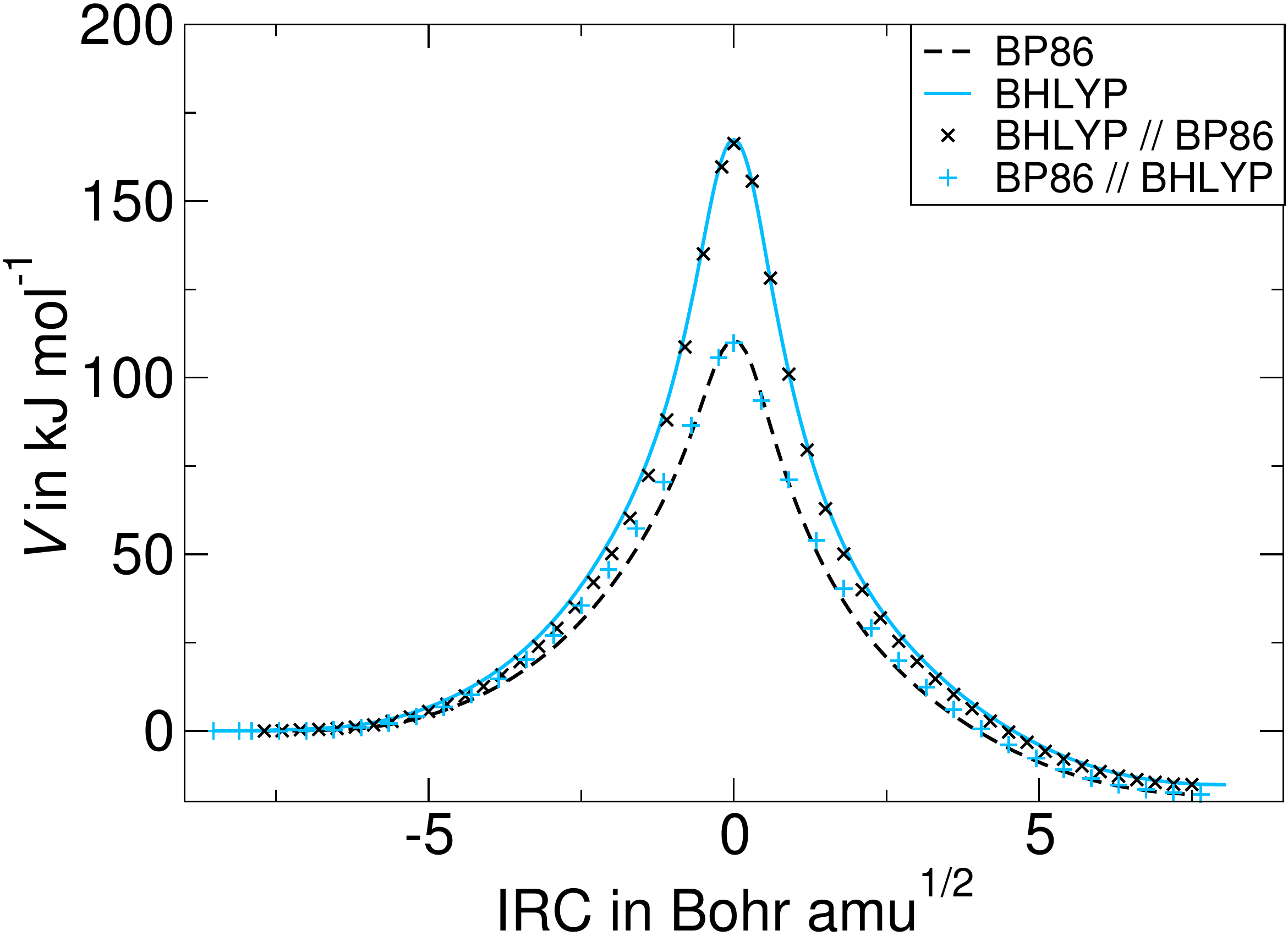}\\
    \includegraphics[width=8cm]{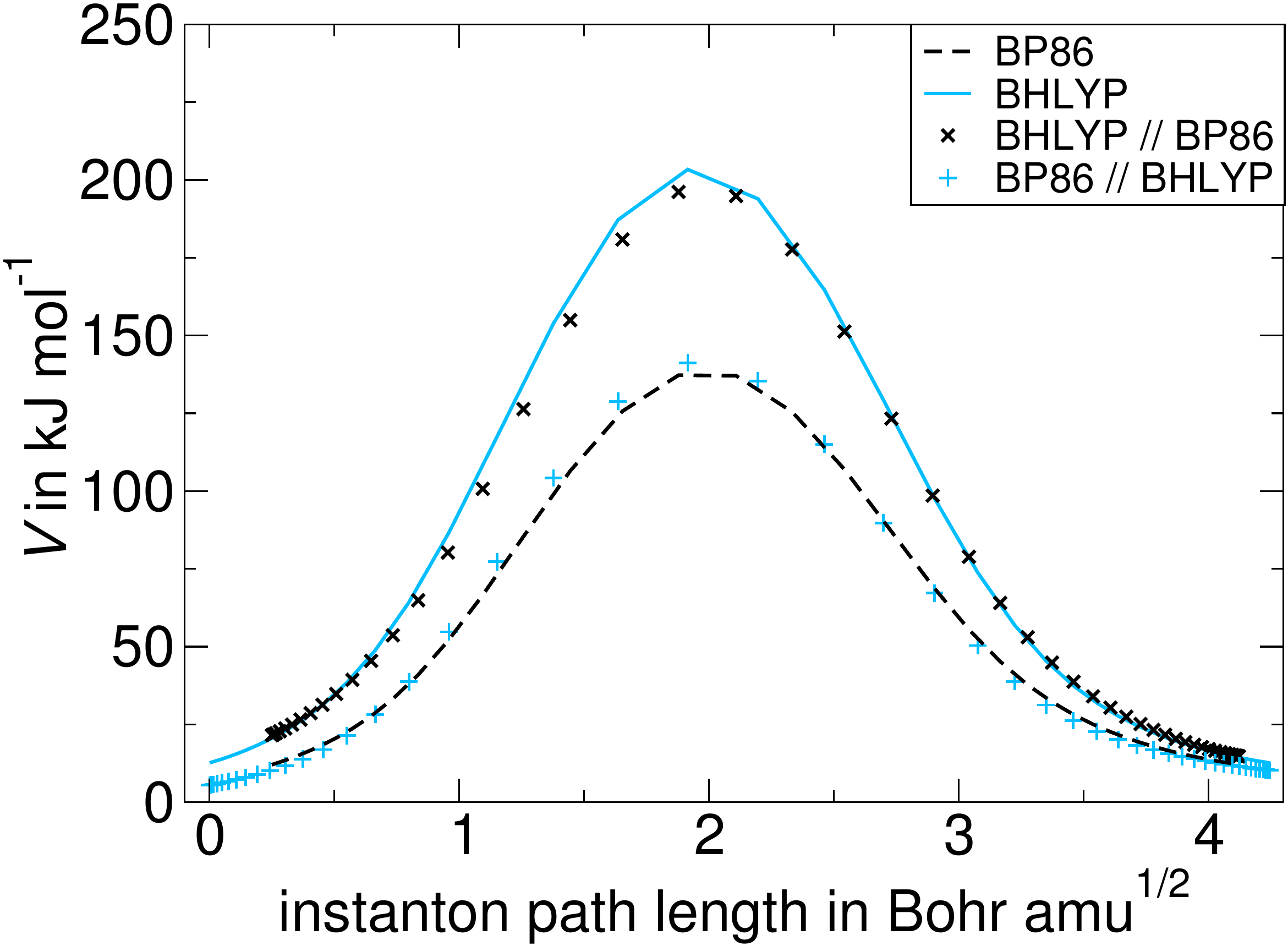}
  \caption{Potential energy along the IRCs of reaction 2 \new{(top) and along
    instanton paths at 100K (bottom)}.
        \label{fig:dual-level:sigmatropicIRC}
  }
\end{figure}

This is a much more severe test for the dual-level, because the potential
activation energy obtained by BHLYP is more than 56.8~kJ~mol$^{-1}$ higher
than the one obtained by BP86\new{, see \tabref{tab:dual-level:sigmatropic}}.
This is attributed to the high amount (50\%) of exact exchange in the BHLYP
functional. \new{CCSD(T)-F12a calculations on the BHLYP (BP86)
  geometries resulted in $V_A=149.1$ kJ mol$^{-1}$ (148.9 kJ
  mol$^{-1}$) and $\Delta V= -11.3$ kJ mol$^{-1}$ ($-11.4$ kJ mol$^{-1}$) implying that the BHLYP/6-31G* method yields more reliable results than the BP86/6-31G* method. }
 However, the two functionals are eminently
suitable to demonstrate the applicability of the dual-level method in cases
where the activation barrier of the two functionals differ significantly.  In
this context, it has to be mentioned that, although the activation barriers
obtained by the two functionals disagree, the reaction energies agree within
2.8~kJ~mol$^{-1}$, see \tabref{tab:dual-level:sigmatropic}.  For both
dual-level combinations, BHLYP /\!/ BP86 and BP86 /\!/ BHLYP,
the potential energy along the IRC is restored quite well, as can be seen in
\figref{fig:dual-level:sigmatropicIRC}. \new{The bottom graph of
  \figref{fig:dual-level:sigmatropicIRC} shows the potential energy along 
  instantons at 100K. The energies are, again, well reproduced by the
  dual-level approach. The lengths of the instantons differ
  somewhat. Comparison of the upper and the lower graph additionally
  shows that the instanton path leads to regions in configuration space with
  significantly higher potential energy than the classical transition
  state (203.4 vs 167.4 kJ mol$^{-1}$ for BHLYP). This is caused by corner-cutting.}

\begin{figure}
    \includegraphics[width=8cm]{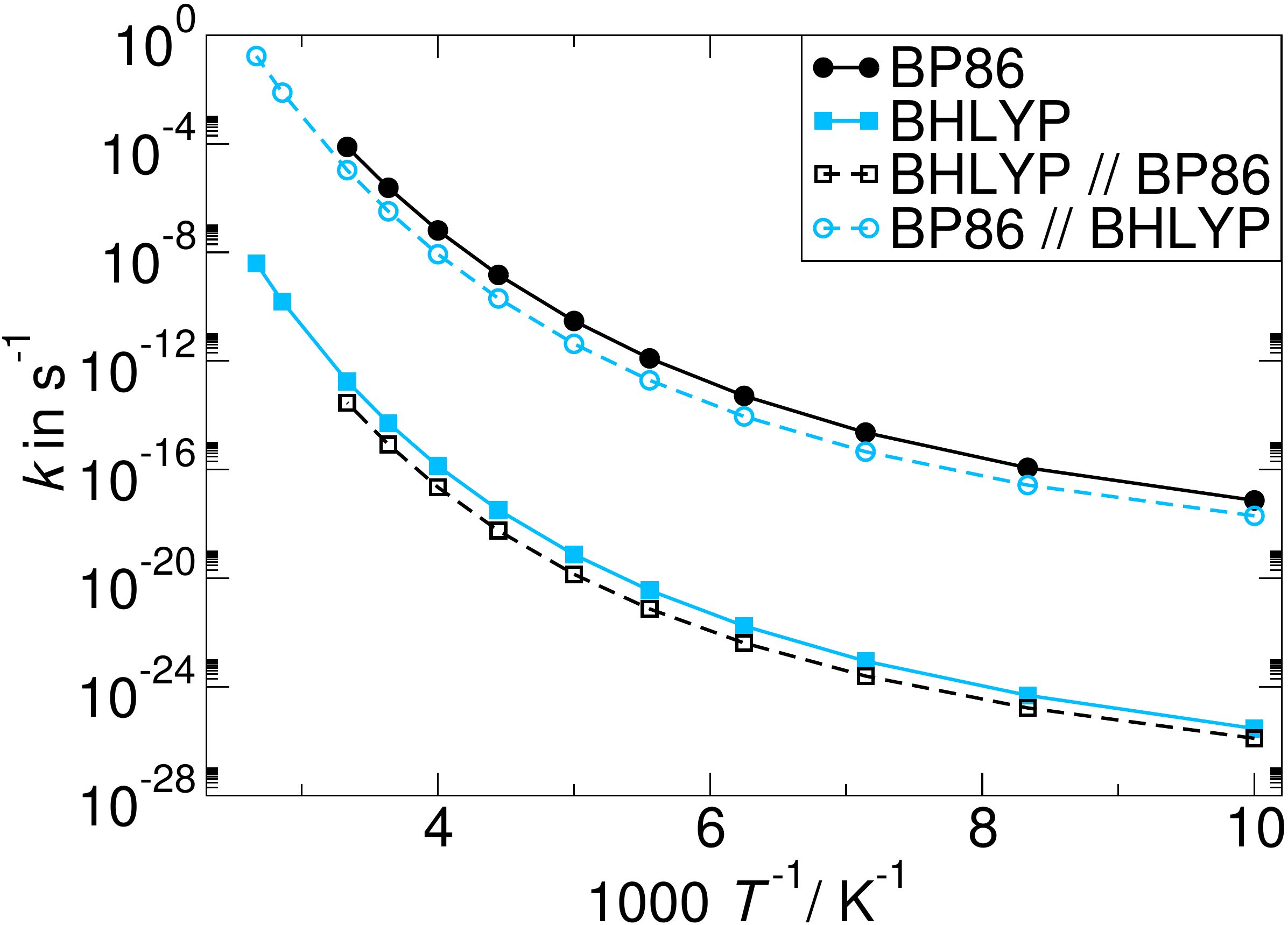}
  \caption{Rate constants for reaction 2.
        \label{fig:dual-level:sigmatropic_rates}
  }
\end{figure}

Instanton rate constants were calculated for a temperature range from 300~K to
100~K using the two density functionals, as well as both dual-level
combinations, see \figref{fig:dual-level:sigmatropic_rates}.  The lower activation
barrier of the BP86 potential compared to the BHLYP potential leads to much
higher reaction rate constants, by more than eight orders of magnitude,
throughout the whole temperature range.  The curvature of all Arrhenius plots
in \figref{fig:dual-level:sigmatropic}, an indication of the importance of
tunneling, is strikingly similar even though the crossover temperatures differ
by more than 20~\%.  Despite the qualitative difference of both functionals,
the results obtained with the dual-level instanton method reproduce the rate
constants of the refined potential successfully in both cases.  The values
obtained by the dual-level approach agree with the full calculations of the
respective refined potentials within one order of magnitude, see
\tabref{tab:1,5-deviations}. There, the average and maximum deviations
(highest or lowest ratios) among the temperatures studied are
shown. \new{The small deviations are mainly due to error
  compensation. For BHLYP /\!/ BP86, the underestimation of $S_0$
  is counterbalanced in the exponential term by an overestimation of $S_\text{pot}$. The
  remaining error in the exponential dominates the total error, but is
  still counterbalanced by an opposite deviation in the
  pre-exponential factor.}

\begin{table}[h!]
 \caption{
   Ratios of rate constants for reaction 2.
        \label{tab:1,5-deviations}
	}
        \begin{tabular}{lrrr}
\hline
&	$\frac{k_{\text{BP86}}}{k_{\text{BHLYP}}}$  
&	$\frac{k_{\text{BHLYP/\!/BP86}}}{k_{\text{BHLYP}}}$
&	$\frac{k_{\text{BP86/\!/BHLYP}}}{k_{\text{BP86}}}$
\\[3pt]
\hline
Max.&	$4.75\cdot 10^{+8} $& 	  $1.63\cdot 10^{-1} $& 	  $1.34\cdot 10^{-1} $ \\
Avg.&	$3.62\cdot 10^{+8} $& 	  $2.36\cdot 10^{-1} $& 	  $1.72\cdot 10^{-1} $ \\
\hline
\end{tabular}
\end{table}

\subsection{Reaction 3: Bimolecular Reaction NH$_2$ + H$_2 \rightarrow$ NH$_3$
  + H}

Finally, the dual-level approach is tested on the bimolecular hydrogen atom
transfer reaction NH$_2$ + H$_2 \rightarrow$ NH$_3$ + H.  This five-atomic
system with eleven electrons is small enough to be easily handled fully on the
CCSD(T)-F12/cc-pVDZ-F12 level using conventional instanton theory as refined
potential.
BHLYP/def2-SVP was used as basic potential.

\begin{table*}[htbp!]
 \caption{
        Potential energy barriers $V_A$, potential reaction energies $\Delta V$ and the
        respective values corrected by zero-point energy, $E_A$ and $\Delta E$
        for reaction 3.
        Energies are in kJ~mol$^{-1}$, crossover temperatures $T_{\text{c}}$  in K.
	\label{tab:dual-level:nh4}
      }
        \begin{tabular}{lrrrrrr}
\hline
& Basic Potential & Refined Potential & Dual-Level \\
\hline
Method			&BHLYP		&  	CCSD(T)-F12	&					\\
Basis set		&def2-SVP       &      	cc-pVDZ-F12     &					\\
        $V_A$           &35.5		&       	41.5		&	40.7				\\
        $\Delta V$      &$-15.1$	&       	$-22.0$		&	$-23.2$				\\
        $E_A    $       &44.2		&       	48.8		&	49.3				\\
        $\Delta E$      &$-0.8$		&      	$-8.0$		&	$-9.0$				\\
        $T_\text{c}$    &355.9		&      	355.8		&					\\
\hline
\end{tabular}
\end{table*}



\begin{figure}[h!]
    \includegraphics[width=8cm]{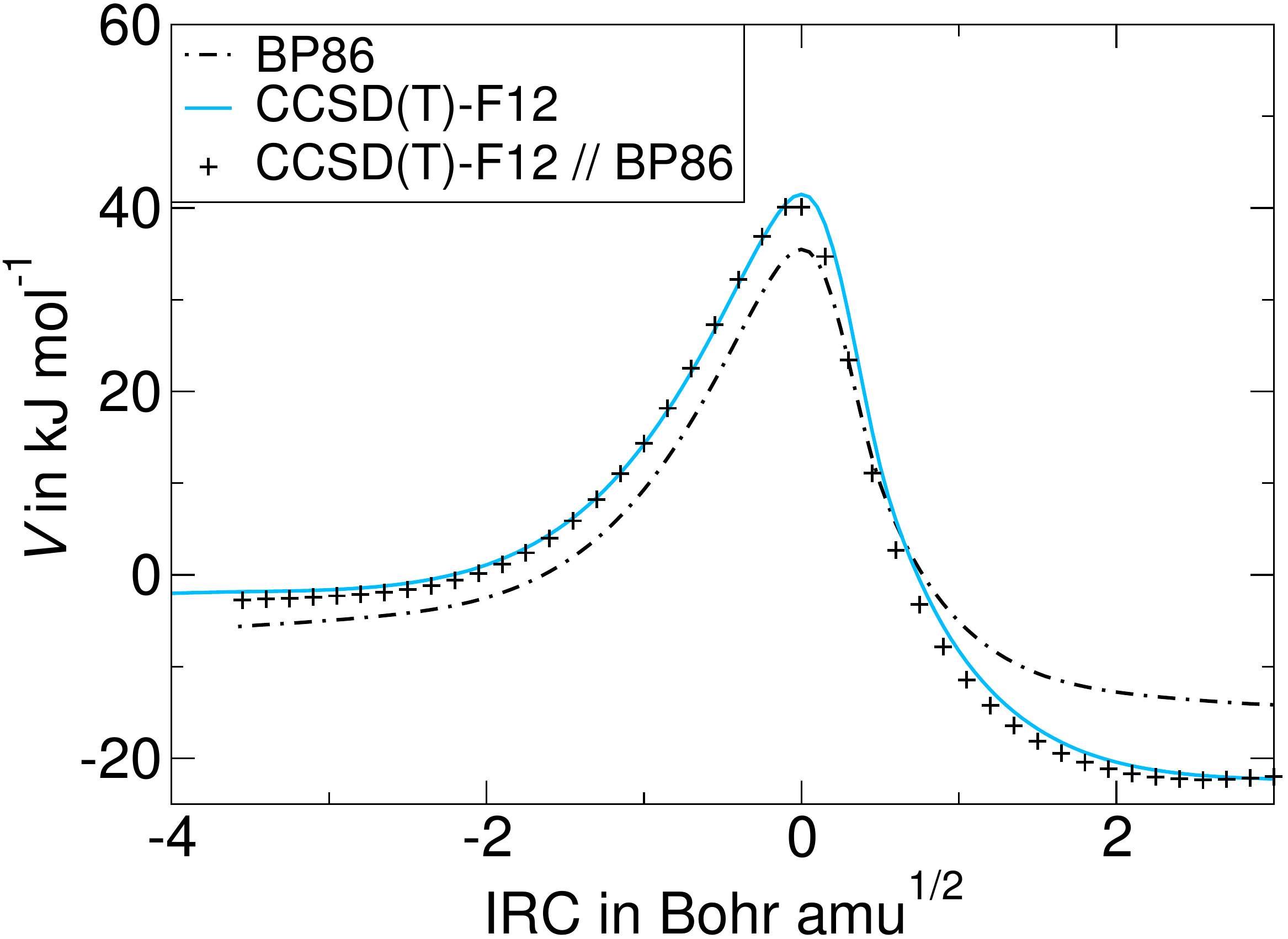}
  \caption{Potential energy along the IRCs of reaction 3 relative to
    the energy of the separated reactants.
    \label{fig:dual-level:nh4IRC}
  }
\end{figure}

On CCSD(T)-F12 level, the electronic reaction energy is $-22.0$~kJ~mol$^{-1}$,
while the vibrationally adiabatic reaction energy, i.e. the electronic energy
plus zero-point vibrational energy, is $-8.0$~kJ~mol$^{-1}$.  During the
reaction, an N--H bond is formed and a H--H bond is broken, leading to a
large difference in zero-point vibrational energy between the reactants and
products.  The BP86 functional underestimates the exothermicity with and
without zero-point energies.  Thus, following from the results obtained for
the Eckart barrier we would expect this case to be rather challenging for the
dual-level approach.
In the region of the transition structure, the CCSD(T)-F12 energy calculations
along the IRC calculated on the BP86 geometries resemble the energy along the
CCSD(T)-F12-IRC well enough, however, see \figref{fig:dual-level:nh4IRC}.

%
%

\begin{figure}[h!]
    \includegraphics[width=8cm]{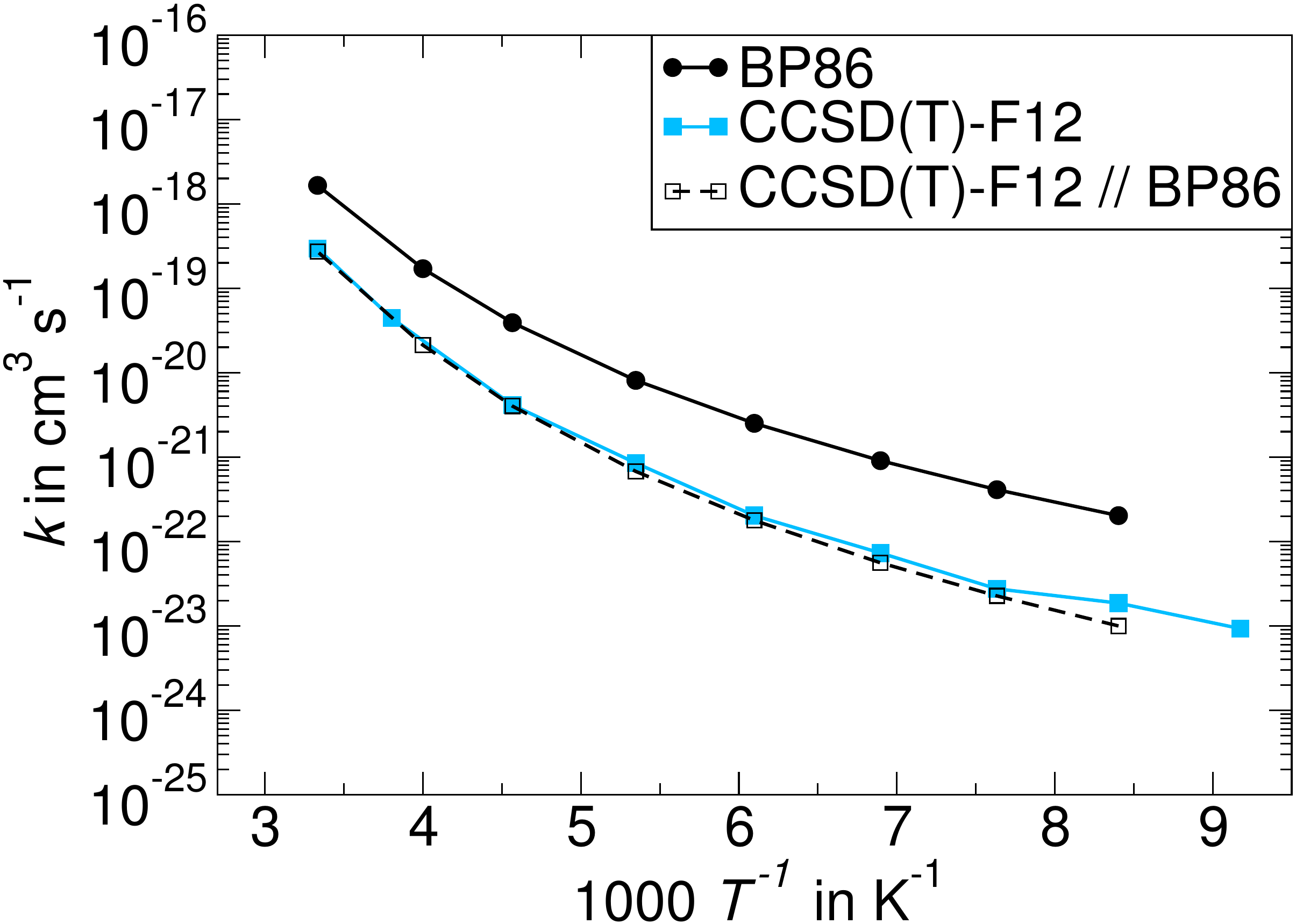}
  \caption{
	Rate constants for reaction 3.
	\label{fig:dual-level:nh4}
  }
\end{figure}

The rate constants obtained with conventional instanton theory
and the corresponding dual-level results are shown in 
\figref{fig:dual-level:nh4}.
BP86 leads to higher reaction rate constants by a factor of 9--12
than the CCSD(T)-F12 reference
throughout the temperature range of 300--110~K, caused by the lower activation barrier.
The dual-level method successfully leads to 
reaction rate constants deviating only by  a factor of less  than 2 from the
results obtained by conventional instanton theory using CCSD(T)-F12.


This finding is quite surprising as we would have expected the 
dual-level approach to fail due to the underestimated exothermicity.
To understand that, let us recall the case of the Eckart barrier (\figref{fig:dual:va}).
There, the potential energy is wrongly described on the right side of the potential energy barrier and
affects the region of the instantons, leading to an incorrect shape and in particular image distribution.
This caused the dual-level approach to be unsuitable.
In the case of reaction 3, the final reaction energy is also described
incorrectly. However, the instanton is restricted to the region in space where
the energy is higher than the energy of the reactants, here to $s <   +0.4$.
There, the basic potential is sufficiently accurate
and the shape and image distribution of the instantons is not affected.
Thus, the dual-level approach is successful in this case.

\section{Conclusion}
We demonstrated the applicability of a dual-level approach to instanton theory
by studying one analytic and three molecular systems.
For the dual-level instanton calculations, all optimizations and Hessian
calculations are carried out using a basic potential, which is computationally
fast, but can be rather approximate.
Additionally, energy calculations only are carried out for all images using a
more accurate refined potential.
We have shown that this approach corrects for the largest part of the errors
in the rate constants by improving the value of $S_{\text{pot}}$ in
\eqref{eq:instantondual-level}.  The instanton geometry
$\vecnp{y}_{\text{inst}}$, the distribution of the images, and therefore the
value of $S_{\text{0}}$ (see \eqref{eq:Snull}) are not changed and assumed to
be described well enough by the basic potential.

The dual-level approach performs astonishingly well in the cases
examined above. The calculated rate constants are similar to the rate
constants obtained by conventional instanton calculations performed on the
refined potential.  For the isomerization reaction of HNC to HCN, the error
with respect to full CCSD(T)-F12 results is reduced from factors of 50--1000 when using
B3LYP to 2.0--2.5 for the CCSD(T)-F12 /\!/ B3LYP dual-level instanton approach.

The method works well even in the case when the activation barrier of the
basic potential, one of the most crucial parameters when considering reaction
rate constants, differs by more than 40\% from the one obtained with the
refined potential, like in reaction 2.  There, the dual-level
method reproduces the corresponding full instanton rate constants within one
order of magnitude while without the dual-level scheme they differ by more
than eight orders of magnitude.


In cases where the exothermicity of the reaction is underestimated, the
dual-level method suffers from a wrong image distribution and hence can not
correct for the errors in the potential energy, as seen from the Eckart
barrier.  However, this problem can easily be probed by performing reference
calculations for the reaction energy.  Furthermore it could be shown for the
reaction of NH$_2$ and H$_2$ that the wrong description of the reaction energy
is problematic only if the region of the instantons is affected. Otherwise,
the dual-level approach performs well. \new{Obviously, there may be
  other causes of errors like a wrongly predicted topology or
  symmetry of the stationary points predicted by the basic potential which we could not identify in the cases we tested.}

In summary, the dual-level approach can help improving the quality of
instanton rate calculations when the systems are too big for the full
treatment with highly accurate electronic structure methods.  It is a
legitimate improvement as long as the basic potential describes the chemical
reaction qualitatively correct.

\section*{Acknowledgments}
This work was financially supported by the German Research Foundation (DFG)
within the Cluster of Excellence in Simulation Technology (EXC 310/2) at the
University of Stuttgart.
This work was
financially supported by the European Union's Horizon 2020 research and
innovation programme (grant agreement No. 646717, TUNNELCHEM).

\bibliography{mod,local_jk}

\clearpage
\newpage
\section*{TOC Graphic}

\includegraphics[width=8.9cm]{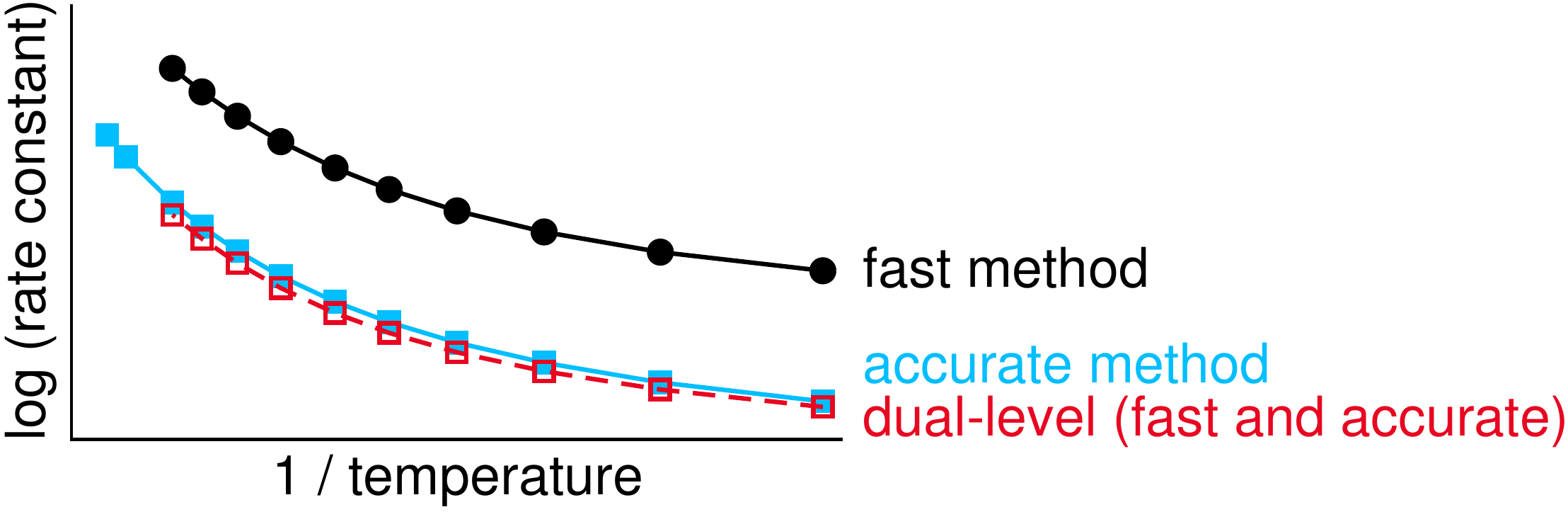}

\end{document}